# Unified constitutive model of body-centered cubic metals informed by discrete dislocation dynamics

Jiarui Zhang[a], Yin Zhang[a,b,*]

[a] State Key Laboratory for Turbulence and Complex Systems, Department of Mechanics, School of Mechanics and Engineering Science, Peking University, Beijing 100871, China

[b] HEDPS, CAPT and IFSA, Collaborative Innovation Center of MoE, Peking University, Beijing 100871, China

**Abstract**

Predicting plastic flow in body-centered cubic metals across temperature and strain-rate regimes requires distinguishing structural resistance from kinetic contributions. Here, we develop a temperature dependent mobility law for screw dislocation, integrating thermal activation theory and relativistic effects. Through discrete dislocation dynamics simulations, uniaxial tension and compression simulations were conducted on configurations with different initial dislocation densities, elucidating the correlation between plastic strain rate and key parameters such as stress, temperature, and dislocation density. Subsequently, a unified constitutive model incorporating temperature and strain rate is derived based on microscopic mechanisms. The model naturally bridges microscopic mechanisms to continuum response, resolving the additivity dilemma by establishing explicit physical criteria for how structural resistances and competing kinetic mechanisms superimpose. Through rigorous bottom-up simulations, we quantitatively capture the crossover point where phonon drag overtakes thermal activation, providing a mechanism-based explanation of anomalous thermal hardening. This work enables physically consistent predictions of strength and creep behavior across broad temperature and strain-rate regimes, advancing beyond phenomenological descriptions toward transparent, mechanism-based alloy design.

*Corresponding author email: yinzhang@pku.edu.cn

## 1. Introduction

Predicting plastic flow in metals across temperatures and strain rates regimes requires constitutive laws that capture changes in the processes controlling deformation (Curtin and Ronald, 2003; Kocks and Mecking, 2003; Roters et al., 2010). Dislocation motion links atomistic processes to macroscopic strength, hardening and creep, but the relevant contributions of deformation mechanisms do not necessarily enter a constitutive law at the same physical level (Bulatov et al., 2006; El-Awady, 2015; Hull and Bacon, 2011). Some mechanisms restrict the mechanical driving force available for dislocation glide, whereas others provide distinct kinetic pathways for generating plastic strain (Kocks et al., 1975). The key modelling question is therefore not simply how to add more mechanisms, but which contributions should be combined as effective resistance stresses and which should be represented as competing plastic-rate contributions.

Refractory body-centered cubic (BCC) metals, such as tungsten (W), serve as ideal model materials to address this multiscale challenge (Cereceda et al., 2016; Li et al., 2021; Srivastava et al., 2020). Unlike low lattice resistance for dislocation glide (Po et al., 2016) and linear drag model (Gilbert et al., 2011) in face-centered cubic metals, the plastic flow in BCC metals is characterized by a striking disparity between the mobility of edge and screw dislocations. It is well-documented in classic dislocation theory that edge dislocations possess low lattice resistance and move rapidly even at low stresses. In contrast, the plastic flow is rate-limited by the compact, non-planar cores of <111> screw dislocations (Po et al., 2016; Starikov and Tseplyaev, 2020). These screw dislocations experience significant lattice friction (Peierls barrier) and display strong strain-rate and temperature-dependent behavior through thermally activated kink-pair formation and migration (Narayanan et al., 2014). Furthermore, the complex collective behavior of dislocation plasticity arises from the interplay of multiple competing deformation mechanisms in addition to the high intrinsic lattice resistance for screw dislocations. These include: (1) phonon drag, arising from interactions between moving dislocations and lattice vibrations, and dominates when dislocation velocities approach orders of magnitude of $10^2$-$10^3$ m/s; (2) forest hardening, reflecting the collective hindrance from dislocation interactions; and (3) relativistic effects, which become significant as dislocation velocities approach the shear wave speed (approx. $3\times10^3$ m/s for W). These features make BCC W a useful model system for separating structural resistance effects from competition between kinetic pathways.

Substantial progress has already been made in describing the individual ingredients of BCC plasticity. Theoretical models have provided valuable physical insights into stress-dependent slip activation energies (Nabarro, 1997), while molecular simulations have revealed the nucleation and migration mechanisms of screw dislocation kink pairs (Chang et al., 2002; Domain and Monnet, 2005; Kuksin and Yanilkin, 2013; Narayanan et al., 2014; Stukowski et al., 2015). Atomistically informed mobility relations for BCC metals have described the transition from thermally activated motion to a viscous regime (Gilbert et al., 2011), while kinetic Monte Carlo studies have quantified thermally activated screw motion in W (Stukowski et al., 2015). Discrete dislocation dynamics (DDD) simulations serve as a bridge between the atomic and continuum scales, enabling the study of collective dislocation evolution under complex loading conditions (Arsenlis et al., 2007; Bulatov and Cai, 2006; Devincre et al., 2011; Ghoniem et al., 2000; Giessen and Needleman, 1995; Kubin et al., 1992; Zbib and Diaz de la Rubia, 2002). Generalized BCC mobility laws have also been formulated and implemented in DDD simulations of W (Po et al., 2016), and multiscale dislocation-dynamics studies have examined temperature and high-rate plasticity in BCC metals (El Ters and Shehadeh, 2019). DDD-based analyses have further connected dislocation density, mobility, strain rate and strength, providing a foundation for density–rate descriptions of plasticity (El-Awady, 2015; Fan et al., 2021). Recent high-rate experiments have highlighted the change from conventional thermal softening to thermal hardening (Borasi et al., 2026; Borasi and Schuh, 2025; Dowding and Schuh, 2024; Dowding and Schuh, 2026). The unresolved issue addressed here is therefore narrower: when structural resistance and competing kinetic mechanisms coexist, which contributions should be combined at the effective-stress level, and which should be combined at the plastic-rate level. To resolve this question, the theoretical framework presented herein distinguishes between the serial superposition of structural resistances governing the effective driving stress and the parallel additive decomposition of the total strain rate across competing kinetic mechanisms.

In this study, as illustrated in Fig. 1, we formulate a temperature-dependent mobility law for BCC W that includes forward and reverse thermally activated transitions, a smooth connection to phonon drag controlled motion and the high-velocity correction (Argon, 2007). We implement the law in a three-dimensional DDD framework and vary temperature, applied stress and initial dislocation density. Using the Orowan relation and the simulated dislocation velocities and densities, we construct a unified constitutive representation for the plastic strain rate and critical resistance. The simulations resolve a density-dependent transition from dislocation starvation to forest hardening and provide a quantitative test of the separation between resistance-level and rate-level contributions. The framework also captures the crossover associated with the change from thermal softening to anomalous thermal hardening. The results demonstrate that the proposed model reproduces the transition from thermally activated to drag-controlled dislocation motion and provides a theoretical foundation for a unified description of microscopic dislocation dynamics and macroscopic plastic response.

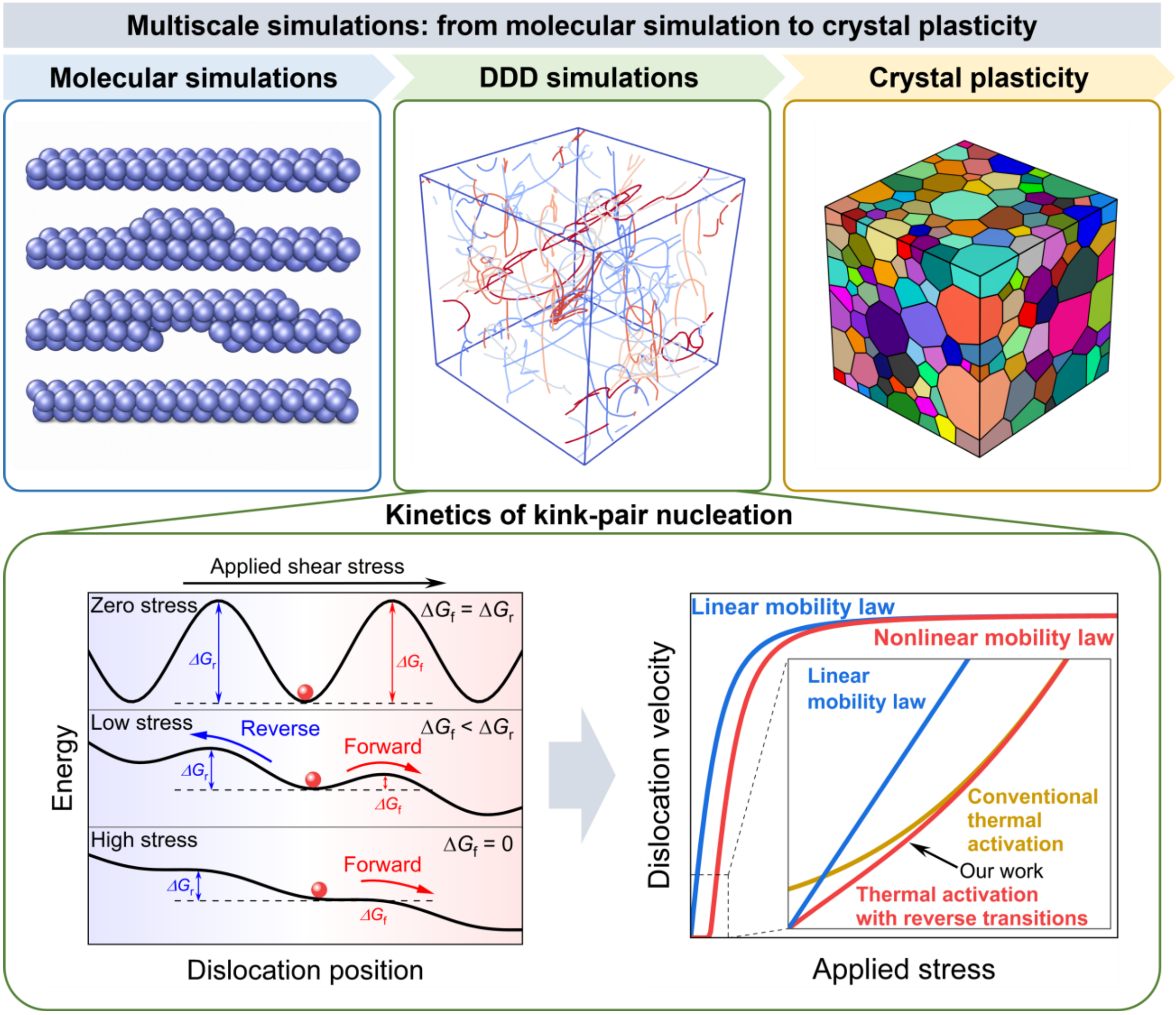

Fig. 1. Schematic of thermally activated dislocation glide mechanisms. Based on molecular simulations, the modified dislocation mobility model incorporates both forward and reverse glide, and has been implemented within the dislocation dynamics simulation framework.

## 2. Temperature dependent dislocation mobility model

The glide velocity of dislocations is influenced by factors such as applied shear stress, crystal purity, temperature, and dislocation type (Ashby and Jones, 2012; Gurrutxaga-Lerma et al., 2020; Hull and Bacon, 2011). Numerous experimental studies have established the relationship between dislocation mobility and stress/temperature in pure metals and alloys (Johnston and Gilman, 1959; Stein and Low, 1960). The general correlation between applied resolved shear stress and average dislocation velocity can be fitted using empirical formula (Gurrutxaga-Lerma et al., 2020)

$$v = A\tau^{k}, \tag{1}$$

where $v$ represents the average dislocation velocity, $\tau$ denotes the applied resolved shear stress, $A$ is a material-dependent constant, and $k$ is the power-law exponent.

Although experimental results exhibit significant variations across different materials and temperatures, it is widely accepted that the relationship between dislocation velocity and shear stress can be divided into three distinct regimes (Gurrutxaga-Lerma et al., 2020): (i) At low applied stresses, the power-law exponent $k > 1$, a phenomenon governed by thermal activation processes; (ii) As stress increases, the exponent $k$ progressively decreases. When the applied resolved shear stress surpasses the Peierls potential barrier $\tau_{\mathrm{c}}$, the system enters a quasi-linear regime characterized by $k \approx 1$, a behavior commonly described as the viscous drag mechanism. For these conditions, Eq. (1) can be reformulated as:

$$Bv = \tau b, \tag{2}$$

where $B$ is the drag coefficient and $b$ denotes the Burgers vector. Within the applicable velocity range, $B$ is primarily governed by lattice vibration (phonon scattering) (Argon, 2007). For cryogenic conditions, the reduced phonon population leads to a corresponding decrease in $B$. It is generally established that $B$ exhibits a linear proportionality with temperature $T$ (Al'shitz and Indenbom, 1975; Argon, 2007; Kocks et al., 1975); (iii) For high stress scenarios where the dislocation velocity approaches the shear wave speed $c_{\mathrm{s}} = \sqrt{\mu/\rho_{\mathrm{m}}}$, with $\mu$ being the shear modulus and $\rho_{\mathrm{m}}$ the material density, the exponent $k < 1$, suggesting the predominance of relativistic effects.

For low applied resolved shear stress conditions ( $\tau < \tau_c$ ), the stress magnitude is insufficient to overcome the thermodynamic energy barrier for dislocation glide. Consequently, dislocations rely on thermal activation to initiate the slip process. In this regime, the dislocation velocity can be described by transition state theory and expressed as (Argon, 2007)

$$v = \nu_0 b \exp\left( -\frac{\Delta G(\tau, T)}{k_B T} \right), \tag{3}$$

where $\nu_0$ represents the attempt frequency, $\Delta G$ denotes the activation energy, and $k_B$ is the Boltzmann constant. The attempt frequency $\nu_0$ for dislocation lines is typically on the order of $10^{11}$ $s^{-1}$ (Cottrell, 2006), as determined by atomic vibrational characteristics. In Eq. (3), the activation energy exhibits dependence on both temperature and shear stress. By invoking the Meyer-Neldel compensation rule, this relationship is explicitly formulated as (Zhu et al., 2008; Zhu et al., 2013)

$$\Delta G(\tau, T) = \Delta H(\tau)\left(1 - \frac{T}{T_m}\right), \tag{4}$$

However, Eq. (3) presents a theoretical inconsistency: when the applied resolved shear stress approaches zero, the predicted velocity becomes negligibly small but remains non-zero. This discrepancy arises because the formulation neglects reverse dislocation motion. While dislocations predominantly glide in the forward direction driven by the Peach-Koehler force, thermally activated reverse jumps against this mechanical driving force can still occur, albeit with a much lower probability. Following transition state theory, the net dislocation glide velocity is determined by the probabilistic bias between forward jumps driven by the applied stress and thermally activated backward jumps. The modified expression should account for the net average glide rate (Argon, 2007)

$$v = \nu_0 b \left[ \exp\left( -\frac{\Delta G_f(\tau, T)}{k_B T} \right) - \exp\left( -\frac{\Delta G_r(\tau, T)}{k_B T} \right) \right], \tag{5}$$

where $\Delta G_{\mathrm{f}}(\tau, T)$ represents the activation energy for forward dislocation motion, and $\Delta G_{\mathrm{r}}(\tau, T)$ denotes the activation energy for reverse motion. In this work, we assume that all such terms adhere to the definition given in Eq. (4).

The energy difference governing these opposing transitions is dictated by the mechanical work done by the local Peach-Koehler force. For $\Delta H_{\mathrm{r}}(\tau)$, the general expression is given by

$$\Delta H_{\mathrm{r}}(\tau) = \Delta H_{\mathrm{f}}(\tau) + C\tau, \tag{6}$$

where the prefactor $C$ represents the characteristic activation volume, approximated as $bl\Delta x$. Here, $l$ denotes the minimum length of an active dislocation segment, and $\Delta x$ signifies the activation distance.

When $\tau \geq \tau_{\mathrm{c}}$, the dislocation velocity becomes governed by phonon scattering, exhibiting linear stress dependence as described by Eq. (2). To ensure smooth transition between the thermally activated regime and phonon-dominated regime, we enforce both function continuity and first-derivative continuity at the transition point, yielding:

$$v = v_0 b\left[1 + \frac{V_{\mathrm{c}}}{k_{\mathrm{B}}T}(\tau - \tau_{\mathrm{c}})\right], \tag{7}$$

where $V_{\mathrm{c}}$ represents the activation volume. According to Eq. (5), the activation volume can be expressed as:

$$V_{\mathrm{c}} = -\left.\frac{\partial \Delta G_{\mathrm{f}}(\tau, T)}{\partial \tau}\right|_{\tau=\tau_{\mathrm{c}}} = -\left.\frac{\partial \Delta H_{\mathrm{f}}(\tau)}{\partial \tau}\right|_{\tau=\tau_{\mathrm{c}}}\left(1 - \frac{T}{T_{\mathrm{m}}}\right), \tag{8}$$

The activation free energy for dislocation glide can be determined through molecular dynamics simulations (Kumagai et al., 2020; Narayanan et al., 2014; Stukowski et al., 2015). The stress dependence of this activation free energy can be empirically described using various functional forms, among which the Kocks formulation has been widely adopted (Kocks et al., 1975)

$$\Delta H_{\mathrm{f}}(\tau) = \Delta H_0\left[1 - \left(\frac{\tau}{\tau_{\mathrm{c}}}\right)^p\right]^q, \tag{9}$$

where $\Delta H_0$ represents the activation enthalpy for dislocation motion at zero shear stress, and $p$ and $q$ are dimensionless fitting parameters.

However, adopting the activation energy formulation in Eq. (9) leads to a physically inconsistent scenario where the activation volume $V_c \approx 0$ at $\tau = \tau_c$. This violates the first-derivative continuity requirement essential for dislocation nucleation modeling and introduces numerical instabilities when integrating activation volumes to reconstruct energy barriers (Faisal and Weinberger, 2019). To resolve this, an alternative formulation was proposed as (Faisal and Weinberger, 2019; Weinberger et al., 2012)

$$\Delta H_f(\tau) = \frac{\Delta H_0}{1 - e^{\hat{\alpha}}}\left[1 - \exp\left(\hat{\alpha}\left(1 - \frac{\tau}{\tau_c}\right)\right)\right]. \tag{10}$$

Although this modified formulation ensures a non-zero activation volume at $\tau = \tau_c$, it achieves lower fitting accuracy compared to Eq. (9). Specifically, while Eq. (9) incorporates two tunable parameters to precisely capture the stress dependence, Eq. (10) relies only on parameter $\hat{\alpha}$, limiting its flexibility in matching experimental data across the full stress range.

Thus, while ensuring a non-zero activation volume, we have modified the form of Eq. (10) as follows

$$\Delta H_f(\tau, T) = \frac{\Delta H_0}{1 - e^{\hat{\alpha}}}\left[1 - \exp\left(\hat{\alpha}\left(1 - \left(\frac{\tau}{\tau_c}\right)^{\hat{\beta}}\right)\right)\right]. \tag{11}$$

For this condition, the activation volume can be expressed as

$$V_c = \frac{\Delta H_0 \hat{\alpha} \hat{\beta}}{\left(e^{\hat{\alpha}} - 1\right)\tau_c}\left(1 - \frac{T}{T_m}\right). \tag{12}$$

Furthermore, considering the thermal softening of the lattice, the critical Peierls stress $\tau_c$ exhibits the temperature dependence, commonly expressed as (Saroukhani and Warner, 2017),

$$\tau_c(T) = \tau_c^0\left(1 - \frac{T}{T_m}\right). \tag{13}$$

Consequently, the phonon scattering coefficient $B$ reduces to a linear function of temperature $T$:

$$B = \frac{k_{\mathrm{B}} T}{V_{\mathrm{c}} \nu_0} = \frac{\left(\mathrm{e}^{\hat{\alpha}} - 1\right) k_B \tau_{\mathrm{c}}^0}{\Delta H_0 \hat{\alpha} \hat{\beta} \nu_0} T. \tag{14}$$

As the stress increases, the dislocation velocity does not grow indefinitely but instead approaches a maximum critical value, the shear wave velocity $c_{\mathrm{s}}$. This behavior is typically described by the following expression (Gillis et al., 1969)

$$v = \frac{\bar{v}}{\sqrt{1 + \frac{\bar{v}^2}{c_{\mathrm{s}}^2}}}. \tag{15}$$

In summary, we obtain the relationship between screw dislocation velocity and shear stress as,

$$\bar{v} = \frac{v}{\sqrt{1 - \frac{v^2}{c_{\mathrm{s}}^2}}} = \begin{cases} \nu_0 b \left[ \exp\left( -\frac{\Delta G_{\mathrm{f}}(\tau, T)}{k_{\mathrm{B}} T} \right) - \exp\left( -\frac{\Delta G_{\mathrm{r}}(\tau, T)}{k_{\mathrm{B}} T} \right) \right] & \tau < \tau_{\mathrm{c}} \\ \nu_0 b \left[ 1 + \frac{V_{\mathrm{c}}}{k_{\mathrm{B}} T} \left( \tau - \tau_{\mathrm{c}} \right) \right] & \tau > \tau_{\mathrm{c}} \end{cases}. \tag{16}$$

In BCC metals, edge dislocations do not migrate through the formation of kink pairs like screw dislocations, but instead move under phonon drag characterized by a linear mobility function. Consequently, their behavior is significantly simpler than that of screw dislocations. By setting $\Delta G = 0$, the ratio of dislocation velocity to shear stress becomes independent of stress and solely temperature dependent. Drawing upon existing molecular dynamics results, we formulate the relationship between edge dislocation velocity and shear stress. Incorporating relativistic effects, the velocity expression for edge dislocations is given by (Po et al., 2016)

$$\bar{v} = \frac{v}{\sqrt{1 - \frac{v^2}{c_{\mathrm{s}}^2}}} = \frac{\tau b}{B_0 + B_1 T}, \tag{17}$$

where $B_0$ and $B_1$ denote the damping coefficients.

In this work, pure W was selected as the representative refractory BCC metal for all simulations. The temperature- and stress-dependent dislocation mobility parameters were rigorously calibrated against published fundamental molecular simulation data (Stukowski et al., 2015), with the comprehensive parameter set detailed in Table 1.

**Table 1** Relevant parameters for dislocation mobility and resistance stress model for W

| $\hat{\alpha}$ | $\hat{\beta}$ | $E_0\,(\mathrm{eV})$ | $\tau_c^0\,(\mathrm{MPa})$ | $\nu_0\,(\mathrm{s}^{-1})$ |
|---|---|---|---|---|
| 2.28 | 1.04 | 1.63 | 1600 | $4.10\times10^{11}$ |
| $b\,(\mathrm{m})$ | $C\,(\mathrm{eV/MPa})$ | $B_0\,(\mathrm{Pa\cdot s})$ | $B_1\,(\mathrm{Pa\cdot s/K})$ | $T_m\,(\mathrm{K})$ |
| $2.74\times10^{-10}$ | $1.19\times10^{-2}$ | $4.26\times10^{-4}$ | $8.70\times10^{-7}$ | 3695.15 |

To ensure the thermodynamic consistency and predictive accuracy of the multiscale framework, the temperature-dependent material properties of pure W (specifically, the mass density, Young's modulus, and Poisson's ratio) are explicitly incorporated. These macroscopic properties are defined as follows (Škoro et al., 2011):

$$\begin{aligned}\rho\left(\mathrm{g/cm^3}\right)&=19.3027-2.3786\times10^{-4}T-2.244\times10^{-8}T^2,\\ E\left(\mathrm{MPa}\right)&=391.448-1.3160\times10^{-2}T-1.4838\times10^{-5}T^2,\\ \nu&=0.279+1.0893\times10^{-5}T.\end{aligned}\tag{18}$$

The atomistic origins of this model are schematically illustrated in Fig. 2(a). To validate this thermodynamic formulation, Fig. 2(b) compares the theoretical activation enthalpy with molecular simulation results. The model accurately captures the stress-dependent energy barrier reduction. Fig. 2(c) presents the glide velocity fitting for screw dislocations, which are intrinsically bottlenecked by high Peierls barriers and heavily rely on thermal activation. Fig. 2(d) illustrates the mobility of edge dislocations; owing to their inherently lower lattice resistance, edge dislocations exhibit significantly higher velocities. The dislocation mobility was fitted based on molecular simulation results.

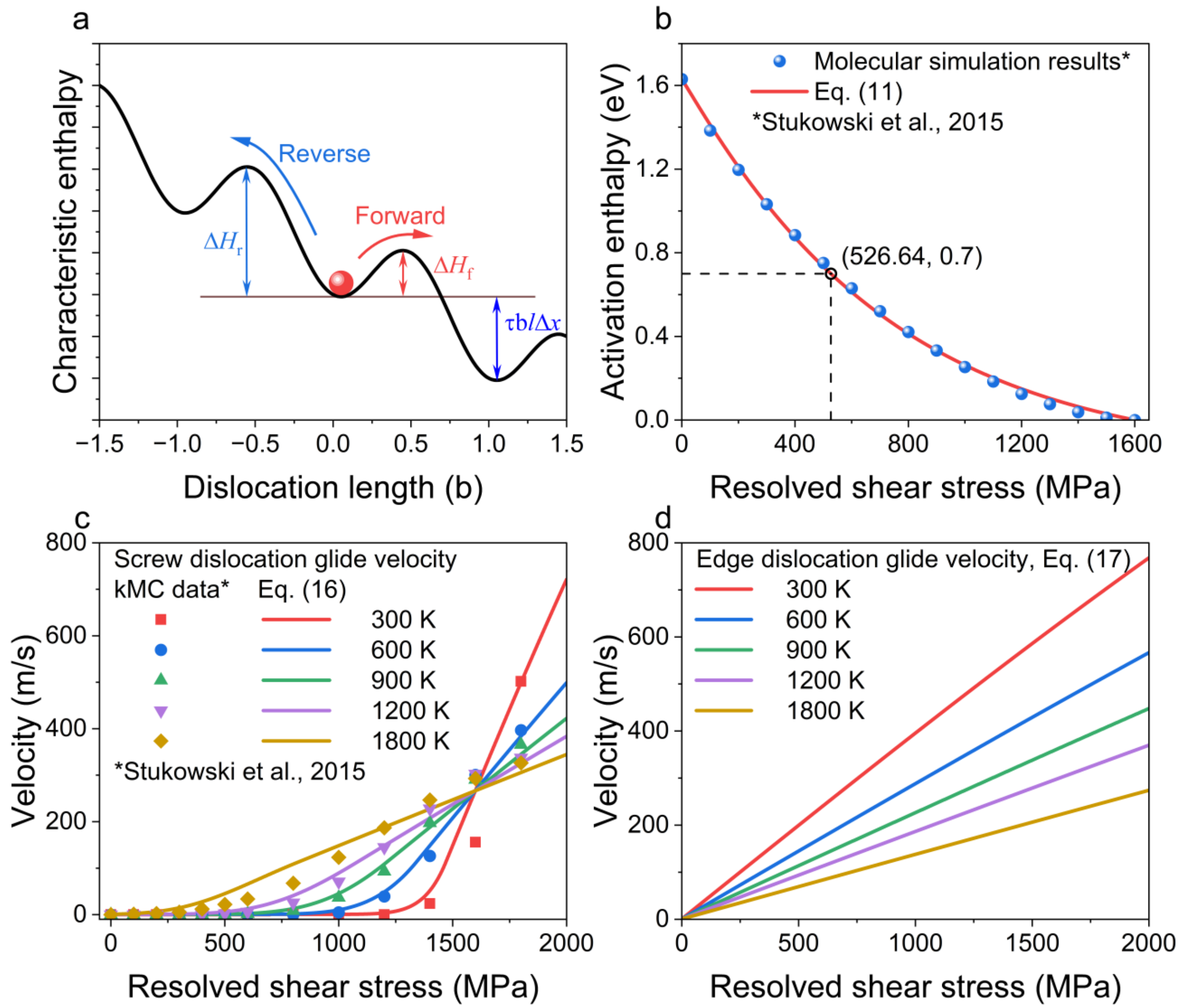

Fig. 2. Temperature-dependent dislocation mobility model. (a) Schematic of thermally activated dislocation glide mechanisms. (b) Comparison between published molecular simulation results and model predictions for activation enthalpy. (c) Fitting of published molecular simulation data for screw dislocation glide velocity using the mobility model. (d) Mobility model for edge dislocation glide velocity.

# 3. Unified strain rate and temperature dependent constitutive model

## *3.1. DDD simulation setup and loading protocol*

In this work, we performed 4,050 DDD simulations using the open-source ParaDiS (v2.7) (Arsenlis et al., 2007), with our modified nonlinear mobility law. The simulations covered temperatures from 300 to 2000 K, initial dislocation densities from $10^7$ to $10^{15}$ m$^{-2}$, and applied stresses from 10 to 8000 MPa. The simulation cell was periodic along the [100], [010], and [001] crystallographic directions. For BCC W, dislocation glide was restricted to the 1/2<111>{110} family. The four unique <111> directions and the three compatible {110} planes per direction give 12 crystallographic slip systems under the convention used here. Dislocation cross-slip was not included in any of the simulations.

The initial configuration comprised straight, infinitely long dislocations distributed uniformly among these slip systems. To reduce periodic-boundary artifacts and approximate bulk behavior, the simulation cell size was scaled with the initial dislocation density according to $L\sqrt{\rho_0} \approx 4.52$ (Fan et al., 2021). Each initial configuration was relaxed without external loading before mechanical loading began. Details of the DDD algorithmic implementation and topological rules, including integration and velocity calculations, network remeshing, and topological operations, are provided in Appendix A.

For each combination of temperature, applied stress and initial dislocation density, six loading cases were initiated from one relaxed dislocation configuration. Uniaxial tension and compression were applied along each of the three orthogonal <100> directions, with the applied stress held constant throughout each case. The resolved shear stress $\tau$ was obtained from the applied uniaxial stress $\sigma$ using $\tau = m_s\sigma$, where the Schmid factor $m_s$ was set to $1/\sqrt{6}$. Plastic strain, plastic strain rate and dislocation density were recorded throughout loading.

The present protocol prescribed stress and measured the resulting plastic strain rate, rather than prescribing strain rate and following the stress response. Within this parameter sweep, it avoided an additional yield-point search and its associated computational expense. This protocol-specific advantage does not establish a general computational-efficiency ranking between constant-stress and constant-strain-rate DDD methods. Transient evolution and the identification of a quasi-steady flow interval remained part of the analysis.

Each loading case was allowed up to 10,000 simulation steps, subject to a maximum computational budget of 96 core hours. Reaching this budget did not imply completion of the prescribed simulation steps. For an initial dislocation density of $1.73\times10^{13}$ m$^{-2}$ at 1200 K, a run at 2000 MPa required approximately 25 core hours. At the same density and temperature, runs below 2000 MPa typically required approximately 3.5 core hours. Many runs above 2000 MPa reached the 96 core hours budget and were terminated. Budget termination was distinct from completion of the prescribed 10,000 simulation steps. These approximate values describe the stated representative conditions, not universal costs. Resource use can vary with stress, temperature, initial dislocation density, network-evolution complexity and the time-step sequence.

*3.2. Steady-state plastic flow and material strength*

The steady-state plastic strain rate served as the characteristic parameter for evaluating the material strength. It was extracted using the maximum-continuous-linear-segment procedure described in Appendix B. A candidate segment was accepted only when $R^2 \geq 0.98$ was satisfied. Runs in which the dislocation network became stationary before this criterion was met were excluded from subsequent analysis. All simulation results are provided in Appendix C, while selected results are presented in Fig. 3.

Fig. 3(a) illustrates the variation of the plastic strain rate $\dot{\varepsilon}^{\mathrm{p}}$ under different temperature and stress conditions. At low stresses (below 1000 MPa), the external stress is insufficient to fully drive screw dislocations over the Peierls barrier athermally. Consequently, the plastic strain rate exhibits pronounced thermal softening, increasing markedly with rising temperature as thermal fluctuations assist kink-pair nucleation. Remarkably, as the applied stress further increases, the strain rate curves for different temperatures converge and intersect at a common point. At this specific crossover, the applied stress equals the athermal resistance, rendering the plastic strain rate completely independent of temperature. Beyond this intersection, at high strain rates (e.g., $10^5$ $s^{-1}$), the plastic strain rate exhibits an approximate power-law dependence on stress, where phonon drag becomes the dominant mechanism impeding dislocation motion. Here, the stress needed to maintain a constant plastic strain rate rises with temperature, leading to a counterintuitive increase in flow stress at higher temperatures. Fig. 3(b) further demonstrates the existence of a friction stress through the equivalent dislocation velocity $\dot{\varepsilon}^{\mathrm{p}}/\rho b$ curve: below a critical stress threshold, the rate remains near zero (indicating immobile dislocations), while beyond this threshold, it rapidly increases.

Fig. 3(c) presents the strain rate response under varying initial dislocation densities at $T = 1200$ K . A higher dislocation density increases the number of mobile dislocations, significantly enhancing the plastic strain rate through the Orowan equation, while simultaneously introducing greater dislocation interaction resistance (i.e., classical forest hardening). In the medium to high stress regime (from 1000 to 4000 MPa), the strain rate curves for different dislocation densities exhibit a tendency to intersect, reflecting a competition between strain rate hardening and forest hardening mechanisms. At even higher stresses (above 4000 MPa), the applied stress is dominated by the internal resistance from forest hardening. In this regime, the strain rate exhibits a consistent power-law dependence on stress, characterized by a unified stress exponent across all dislocation densities. Fig. 3(d) shows the strain rate as a function of dislocation density under different stress conditions at $T = 2000$ K. In lower stress conditions (from 10 to 1000 MPa), once the dislocation density exceeds a critical threshold, further increases in dislocation density exhibit diminishing returns in enhancing the plastic strain rate. This phenomenon indicates a transition to dislocation generation and annihilation dynamics as the rate-limiting mechanism for plastic flow. At high stress (above 1000 MPa), the double-logarithmic plot reveals a slope close to 1, indicating a nearly linear dependence of strain rate on dislocation density. This strong sensitivity of plastic strain rate to low dislocation densities represents a characteristic manifestation of the dislocation starvation effect. A higher dislocation density provides more carriers for plastic deformation while simultaneously enhancing resistance to mutual interactions, which is consistent with the classical forest hardening mechanism.

Building upon these observed trends, the dominant resistance mechanisms can be delineated at different conditions. When the applied resolved shear stress approaches the Peierls stress threshold at relatively low temperatures, dislocation motion is primarily hindered by the intrinsic lattice resistance. In this regime, plastic deformation exhibits strong dependence on temperature and strain rate. At elevated stresses where dislocations enter the high velocity glide regime, phonon scattering emerges as the predominant resistance mechanism. This is characterized by a power-law dependence of strain rate on stress and a negative correlation with temperature. For extremely low dislocation densities, the material demonstrates abrupt strengthening due to dislocation starvation, where available dislocations are rapidly exhausted. In the medium to high stress regime, forest hardening becomes particularly pronounced with increasing dislocation density. A higher dislocation density increases the number of dislocation intersections, thereby elevating the flow stress through enhanced forest hardening.

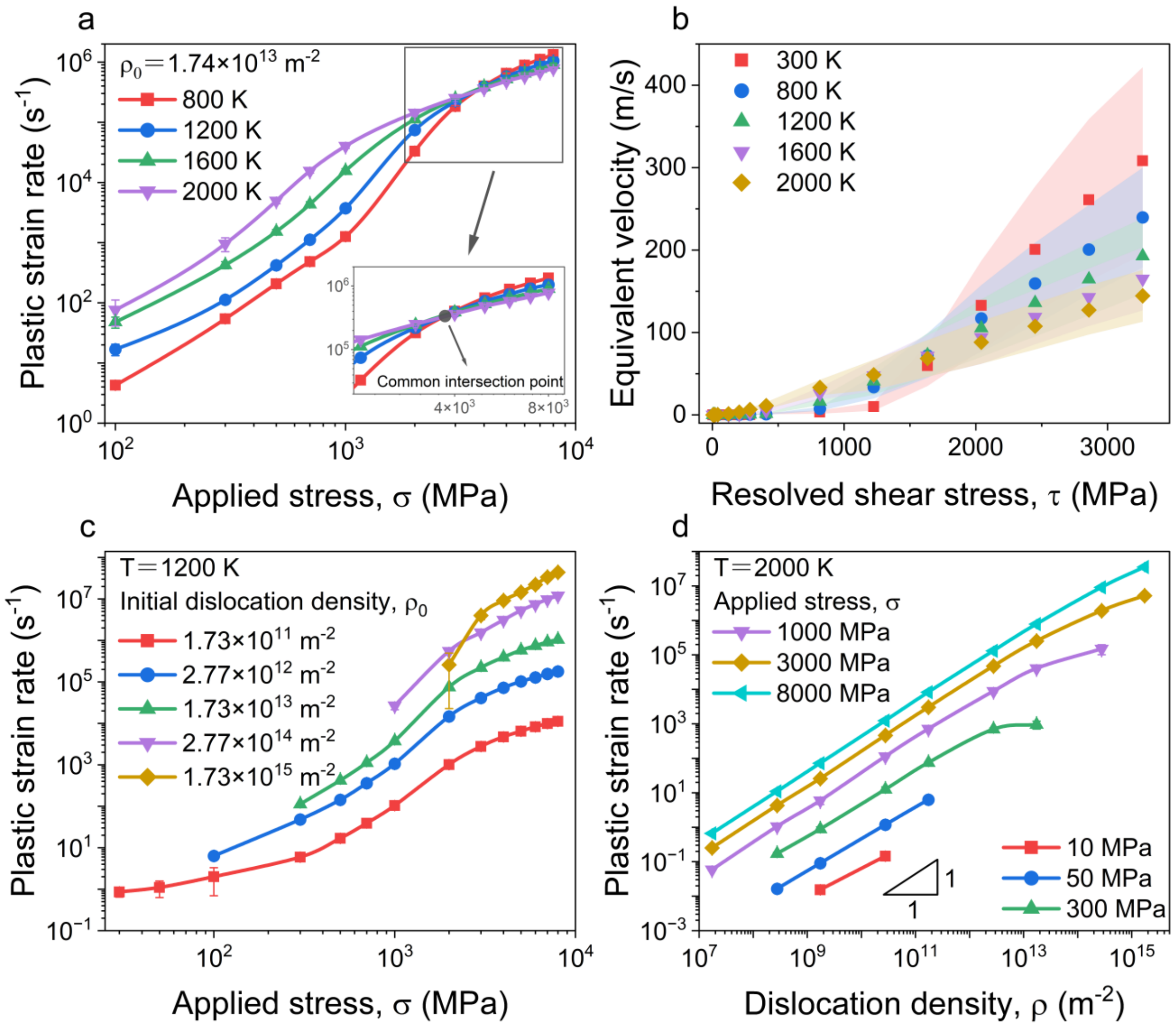

Fig. 3. Selected results from DDD simulations. (a) Stress-dependent plastic strain rate at varying temperatures. (b) Stress-dependent equivalent dislocation glide velocity at different temperatures. (c) Plastic strain rate versus stress for varying dislocation densities. (d) Plastic strain rate versus dislocation density under different applied stresses. Error bars in all figures represent the standard deviation of the six loading-case responses.

### 3.3. Strain rate and temperature dependent plasticity model

To establish the relationship between material strength and plastic strain rate, temperature, and dislocation density, we employ the Orowan equation,

$$\dot{\varepsilon}^{\mathrm{p}} = \rho_{\mathrm{a}} b v, \tag{19}$$

which governs plastic deformation by linking the plastic strain rate to the dislocation density and the average dislocation velocity $v$, $\rho_\mathrm{a}$ denotes the mobile dislocation density of the active slip systems. Our constitutive analysis follows the database-based strain-rate and dislocation-density logic established by Fan et al (Fan et al., 2021). Here, we assume the mobile dislocation density is on the same order of magnitude (Fan et al., 2021) as the total dislocation density $\rho$, i.e. $\rho_\mathrm{a} = f_\mathrm{a}\rho = 2/3\,\rho$.

Considering an additive decomposition of the total plastic strain rate into the thermal activation process and phonon drag mechanism, each governed by its own effective resolved shear stress wherein only resistances operating at scales well separated from the active mechanism are excluded,

$$\dot{\varepsilon}^\mathrm{p} = \dot{\varepsilon}_1^\mathrm{p}\rho_\mathrm{a}b^2\left[\exp\left(-\frac{\Delta\tilde{G}_\mathrm{f}}{k_\mathrm{B}T}\right)-\exp\left(-\frac{\Delta\tilde{G}_\mathrm{r}}{k_\mathrm{B}T}\right)\right]+\dot{\varepsilon}_2^\mathrm{p}\rho_\mathrm{a}b^2\left(\frac{\tilde{V}}{k_\mathrm{B}T}\left(\tau-\tau_0\right)\right)^n\zeta\left(\tau-\tau_0\right). \quad (20)$$

Here, $\Delta\tilde{G}_\mathrm{f}$ and $\Delta\tilde{G}_\mathrm{r}$ denote the apparent activation energies for forward and reverse processes, respectively, which differ from the activation energy associated with the elementary dislocation glide process. $\tilde{V}$ is the apparent activation volume, and $\zeta\left(x\right)$ represents the Heaviside step function. At low stresses, the first term in the equation dominates, reflecting thermally activated behavior. When the applied resolved shear stress exceeds the total resistance, the plastic deformation of the material becomes governed by phonon scattering. The critical resistance can be expressed as

$$\tau_0 = \underbrace{s_\mathrm{e}^0\left(1-\frac{T}{T_\mathrm{m}}\right)}_{\text{Peierls stress, } s_\mathrm{e}} + \underbrace{\alpha\mu b\sqrt{\rho}}_{\text{forest hardening, } \tau_\mathrm{f}} + \underbrace{\beta\mu b\frac{h}{\rho}}_{\text{dislocation starvation, } \tau_\mathrm{s}}, \quad (21)$$

where $\mu$ denotes the shear modulus, while $\alpha$ and $\beta$ are dimensionless parameters, and $h$ is an effective fitted coefficient that controls the magnitude of the starvation contribution (Monavari and Zaiser, 2018; Zaiser et al., 2007). The total resistance originates from three components. The first term is the Peierls stress, $s_\mathrm{e}$, and the second term represents the forest hardening resistance, $\tau_\mathrm{f}$. Furthermore, at low dislocation densities, combining the Orowan equation with the equation of dislocation motion $Bv = \tau b$ yields $\tau = B\dot{\varepsilon}^{\mathrm{p}} / \rho_a b^2$. This establishes that the stress is inversely proportional to the dislocation density, indicating that this third term is attributed to the dislocation starvation mechanism, $\tau_{\mathrm{s}}$.

According to transition state theory, the thermal activation of dislocations is governed by the activation free energy, $\Delta\tilde{G}$, which encompasses both enthalpic and entropic contributions. In the present formulation, the term $\Delta\tilde{H}$ (denoted as $\Delta\tilde{H}_{\mathrm{f}}$ and $\Delta\tilde{H}_{\mathrm{r}}$ for forward and reverse glide) represents the stress-dependent activation enthalpy, at 0 K. This barrier is phenomenologically described as:

$$\Delta\tilde{H}_{\mathrm{f}} = \Delta\tilde{H}_0 \left(1 - \left(\frac{\tau - \tau_{\mathrm{f}} - \tau_{\mathrm{s}}}{s_{\mathrm{e}}}\right)^p\right)^q, \quad \Delta\tilde{H}_{\mathrm{r}} = \Delta\tilde{H}_{\mathrm{f}} + \tilde{C}\left(\tau - \tau_{\mathrm{f}} - \tau_{\mathrm{s}}\right), \tag{22}$$

where $\Delta\tilde{H}_0$ denotes the zero-stress energy barrier, and $\tilde{C}$ is a constant. Specifically, when dislocations overcome the energy barrier for forward glide, the first term in Eq. (20) becomes constant, i.e., when $\tau \geq \tau_0$, $\Delta\tilde{H}_{\mathrm{f}} \equiv 0$. Based on the thermodynamic compensation law (the Meyer-Neldel rule), the activation entropy $\Delta\tilde{S}$ is assumed proportional to the activation enthalpy (Zhu et al., 2008; Zhu et al., 2013). Substituting this entropic relation into the fundamental thermodynamic equation yields the apparent activation energy:

$$\Delta\tilde{G} = \Delta\tilde{H}\left(1 - \frac{T}{T_{\mathrm{m}}}\right). \tag{23}$$

Specifically, Eq. (20) is designed such that when the applied resolved shear stress reaches the temperature-independent athermal resistance

$$\tau_{\mathrm{r}} = s_{\mathrm{e}}^0 + \tau_{\mathrm{f}} + \tau_{\mathrm{s}}, \tag{24}$$

the same dislocation density yields identical plastic strain rates across different temperatures, with the corresponding strain rate being

$$\dot{\varepsilon}_{\mathrm{r}}^{\mathrm{p}} = \left[\dot{\varepsilon}_1^{\mathrm{p}} + \dot{\varepsilon}_2^{\mathrm{p}}\left(\frac{\tilde{V}s_{\mathrm{e}}^0}{k_{\mathrm{B}}T_{\mathrm{m}}}\right)^n\right]\rho_{\mathrm{a}}b^2. \tag{25}$$

This condition corresponds to the common intersection point in Fig. 3(a). By systematically extracting these common intersection points, we obtain the results presented in Fig. 4. Fig. 4(a) displays the functional dependence of critical stress $\tau_{\mathrm{r}}$ on dislocation density, exhibiting a characteristic non-monotonic trend where $\tau_{\mathrm{r}}$ first decreases and then increases with increasing dislocation density. This behavior signifies a fundamental transition in deformation mechanism from dislocation starvation dominance at lower densities to the forest hardening regime at higher densities. Fig. 4(b) reveals a distinct linear relationship between plastic strain rate and dislocation density, confirming the validity of our constitutive model.

Although the full additive model proposed above comprehensively captures the governing mechanisms across the entire stress range, the strong coupling between temperature and stress complicates the direct evaluation of discrete simulation results. To decouple these thermodynamic variables and reveal underlying universal scaling laws, we normalized the data to facilitate further quantitative analysis. Specifically, we partitioned the plastic behavior into two distinct stress regimes.

In the low-stress regime, i.e., when $\tau < \tau_0$, plastic deformation is predominantly governed by thermal activation, implying that the first term in Eq. (20) dominates. This allows the establishment of temperature-independent relationships between stress ratio $P_1$ and normalized strain rate $Q_1$,

$$P_1 = \frac{\tau - \tau_{\mathrm{f}} - \tau_{\mathrm{s}}}{s_{\mathrm{e}}}, \quad Q_1 = \frac{\dot{\varepsilon}^{\mathrm{p}}}{\dot{\varepsilon}_1^{\mathrm{p}} \rho_{\mathrm{a}} b^2}. \tag{26}$$

Both $P_1$ and $Q_1$ are dimensionless parameters bounded between 0 and 1. The collective behavior follows an approximate exponential function of the form

$$Q_1 = \exp\left[-\tilde{A}\left(1 - P_1^p\right)^q\right] - \exp\left[-\tilde{A}\left(1 - P_1^p\right)^q - \tilde{B}P_1\right], \tag{27}$$

where $\tilde{A} = \tilde{E}_0\left(T_{\mathrm{m}} - T\right)/k_{\mathrm{B}}TT_{\mathrm{m}}$, $\tilde{B} = \tilde{C}s_{\mathrm{e}}\left(T_{\mathrm{m}} - T\right)/k_{\mathrm{B}}TT_{\mathrm{m}}$, the corresponding fitting results are presented in Fig. 4(c).

In high-stress conditions ($\tau \geq \tau_0$), we introduce $P_2$ and $Q_2$ as dimensionless stress intensity and strain-rate characterization parameters, respectively

$$P_2 = \frac{\tilde{V}}{k_{\mathrm{B}}T}\left(\tau - \tau_0\right), \quad Q_2 = \frac{\dot{\varepsilon}^{\mathrm{p}}}{\dot{\varepsilon}_2^{\mathrm{p}} \rho_a b^2}, \tag{28}$$

where $P_2$ is a dimensionless effective stress and $Q_2$ denotes the normalized strain rate. Plastic deformation is governed by phonon scattering, where the first term in Eq. (20) reduces to a constant. This yields the following constitutive relationship,

$$Q_2 = P_2^n + \dot{\varepsilon}_1^{\mathrm{p}} / \dot{\varepsilon}_2^{\mathrm{p}}, \tag{29}$$

The plastic response exhibits a robust power-law scaling across all investigated temperatures, as demonstrated by the unified results presented in Fig. 4(d). The complete set of extracted parameters utilized in the proposed strain-rate- and temperature-dependent plasticity model is systematically summarized in Table 2.

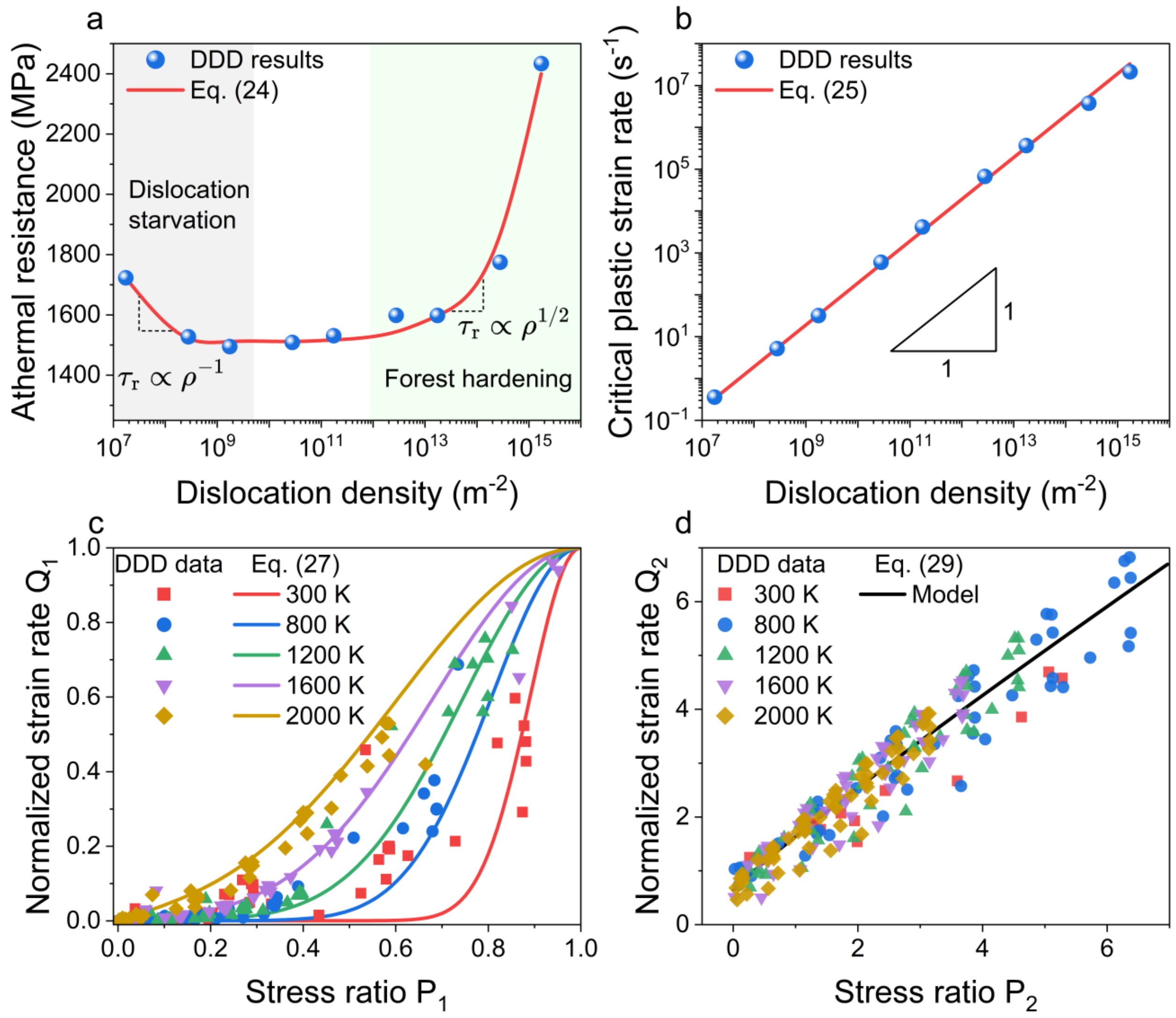

Fig. 4. Scaling relationships and validation of the unified constitutive model. (a) Three distinct regimes governing the dependence of critical stress on dislocation density. (b) Linear scaling between the critical plastic strain rate (corresponding to the intersection point in Fig. 3(a)) and dislocation density. (c, d) DDD simulations with the nonlinear mobility law, comparing the theoretical predictions with numerical results in (c) the thermally activated regime at low stresses and (d) the phonon drag-dominated regime at intermediate to high stresses.

### *3.4. Mechanism map for dislocation plasticity*

Based on the preceding analysis, the dominant mechanisms under varying loading conditions can be summarized, as illustrated in Fig. 5. At extremely low dislocation densities, the scarcity of activatable dislocations due to the dislocation starvation mechanism results in higher material strength. The underlying mechanism is dictated by the Orowan equation $\dot{\varepsilon}^{\mathrm{p}} = \rho_{\mathrm{a}} b v$. To maintain steady plastic flow, the velocity of the few remaining dislocations must increase dramatically, which in turn requires higher applied stresses. This explains the paradoxical strength increase despite a reduced dislocation density. As the dislocation density increases, the strength initially decreases and then rises. This trend arises because an increasing number of mobile dislocations facilitates plastic deformation, whereas at very high dislocation densities, the forest hardening mechanism becomes dominant.

Furthermore, as temperature increases, the thermal activation regime shrinks because elevated temperatures enhance the capability of dislocations to overcome energy barriers. Upon examining the iso-strain-rate contours in the flow stress and dislocation density space, two distinct geometric features emerge. First, at any given temperature, the contour slope exhibits a pronounced transition when shifting from the thermal activation dominated regime to the phonon drag dominated regime. This transition reflects a fundamental shift in the rate-limiting mechanism of dislocation motion: from stress-assisted thermal fluctuations to viscous glide restricted by lattice dissipation. Second, a comparison of the contour slopes across different temperatures demonstrates that the intrinsic slopes within both regimes are highly temperature dependent. Physically, elevated temperatures supply greater thermal energy, thereby exponentially facilitating dislocation bypass of obstacles in the thermal activation dominated regime. An increase in temperature intensifies lattice vibrations, which subsequently increases the viscous drag coefficient in the phonon drag dominated regime. Especially, deformation at lower temperatures exhibits a higher strain-rate sensitivity.

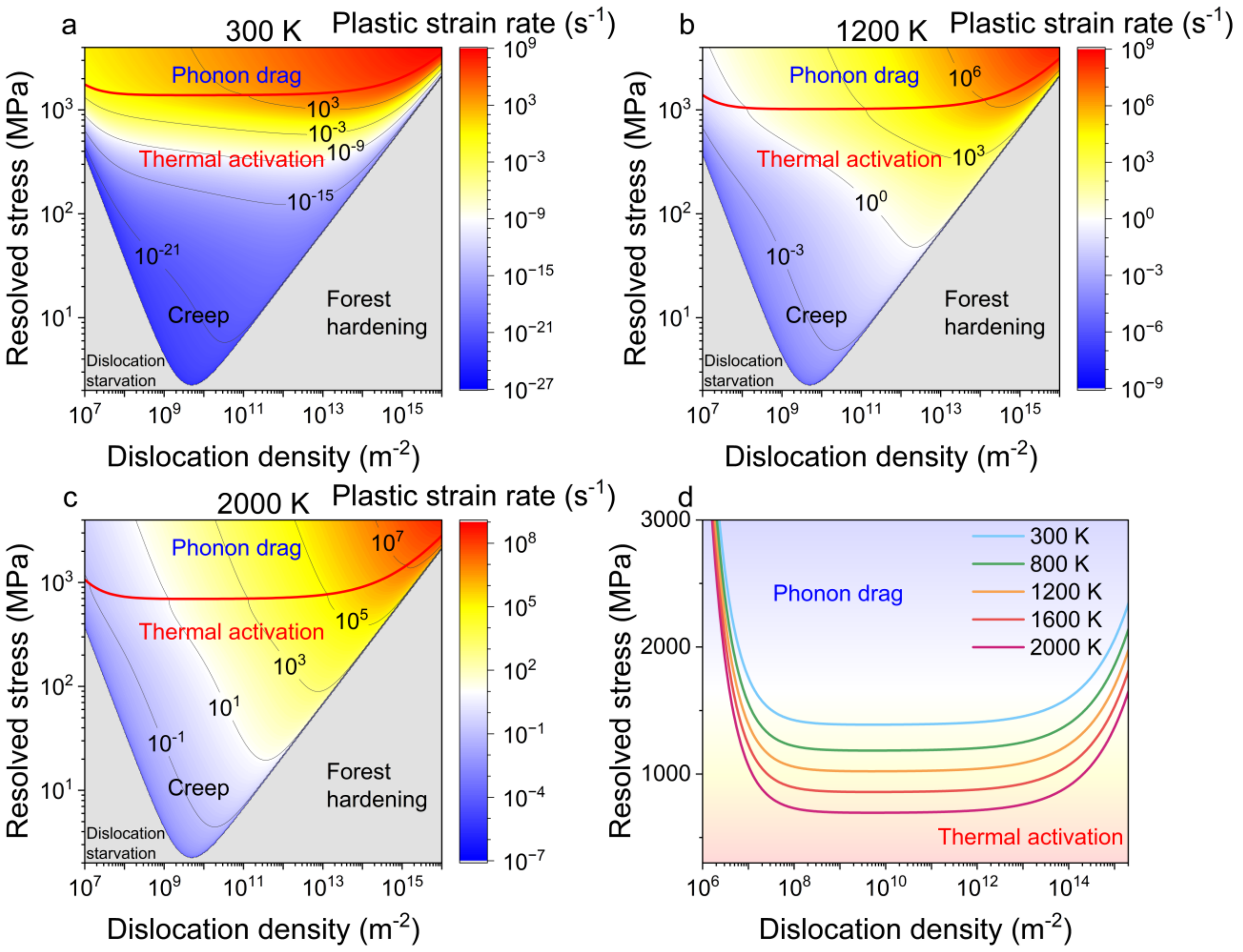

Fig. 5. Mechanism maps for dislocation plasticity. (a)-(c) Maps corresponding to 300 K, 1200 K, and 2000 K, respectively. The solid red lines delineate the transition boundary between the thermally activated regime and the phonon drag regime. (d) Temperature dependence of the transition boundaries between the thermally activated and phonon drag mechanisms.

During the deformation stage, the regions dominated by thermally activated mechanisms and those governed by phonon scattering can be distinguished. At low applied stresses, the behavior can be approximately described as creep. At high stresses, however, the resistance to dislocation glide becomes inversely proportional to temperature, leading to a decrease in plastic strain rate with increasing temperature. This transition corresponds to the shift from thermal softening to thermal hardening mechanisms. These two phenomena will be analyzed in detail in subsequent sections.

*3.5. Low-stress creep behavior*

In this section, we analyze the power-law creep behavior under low-stress conditions. At low applied stresses, the plastic flow rule is governed by the activation energy for screw dislocation motion, as described by transition state theory (Argon, 2007),

$$\dot{\varepsilon}^{\mathrm{p}} = \dot{\varepsilon}_1^{\mathrm{p}} \rho_{\mathrm{a}} b^2 \left[ \exp\left( -\frac{\Delta \tilde{G}_{\mathrm{f}}}{k_{\mathrm{B}} T} \right) - \exp\left( -\frac{\Delta \tilde{G}_{\mathrm{r}}}{k_{\mathrm{B}} T} \right) \right], \tag{30}$$

where $\dot{\varepsilon}_1^{\mathrm{p}}$ is the reference strain rate. The effective resolved shear stress is defined as $\tilde{\tau} = \tau - \tau_{\mathrm{f}} - \tau_{\mathrm{s}}$, excluding the forest hardening resistance and dislocation starvation effect. When $0 < \tilde{\tau} \ll s_{\mathrm{e}}$, the relationship between $\Delta \tilde{G}_{\mathrm{f}}$ and $\Delta \tilde{G}_{\mathrm{r}}$ can be linearly approximated to first order,

$$\Delta \tilde{G}_{\mathrm{f}} \left( \tilde{\tau}, T \right) \cong \Delta \tilde{F} - \tilde{V} \tilde{\tau}, \quad \Delta \tilde{G}_{\mathrm{r}} \left( \tilde{\tau}, T \right) \cong \Delta \tilde{F} + \tilde{V} \tilde{\tau}. \tag{31}$$

Here, $\tilde{V}$ represents the apparent activation volume. $\Delta \tilde{F}$ denotes the Helmholtz free energy at zero stress. According to the Meyer-Neldel compensation rule, this can be expressed as

$$\Delta \tilde{F} = \Delta \tilde{E} \left( 1 - \frac{T}{T_{\mathrm{m}}} \right), \tag{32}$$

where $\Delta \tilde{E}$ represents the activation energy. Based on Eq. (31), Eq. (30) can be simplified to

$$\dot{\varepsilon}^{\mathrm{p}} = \dot{\varepsilon}_1^{\mathrm{p}} \rho_{\mathrm{a}} b^2 \exp\left( -\frac{\Delta \tilde{F}}{k_{\mathrm{B}} T} \right) \sinh\left( \frac{\tilde{V} \tilde{\tau}}{k_{\mathrm{B}} T} \right). \tag{33}$$

At very low stress levels, all inelastic processes exhibit an asymptotically linear stress dependence, expressed as follows:

$$\dot{\varepsilon}^{\mathrm{p}} = \frac{\dot{\varepsilon}_1^{\mathrm{p}} \rho_{\mathrm{a}} b^2 \tilde{V} \tilde{\tau}}{k_{\mathrm{B}} T} \exp\left( -\frac{\Delta \tilde{F}}{k_{\mathrm{B}} T} \right). \tag{34}$$

We selected low-stress regime data for analysis to establish the relationship between stress ratio $P_1$ and normalized strain rate $Q_1$ as well as temperature, denoted as

$$\ln Q_1 = \ln P_1 + \ln \frac{\tilde{V} s_e}{k_B T} - \frac{\Delta \tilde{F}}{k_B T}. \tag{35}$$

To evaluate the creep behavior under low-stress conditions, we first analyzed the data utilizing the dimensionless formulation Eq. (35). As illustrated in Fig. 6(a), the normalized parameters, $Q_1$ and $P_1$, across various temperatures exhibit agreement with the theoretical fitting curve. Furthermore, to directly validate the proposed physical constitutive model Eq. (34), its theoretical predictions were compared with the absolute DDD simulation results, as depicted in Fig. 6(b). The consistency observed in both the dimensionless and physical spaces demonstrates that this formulation effectively characterizes the macroscopic creep behavior governed by dislocation glide at elevated temperatures.

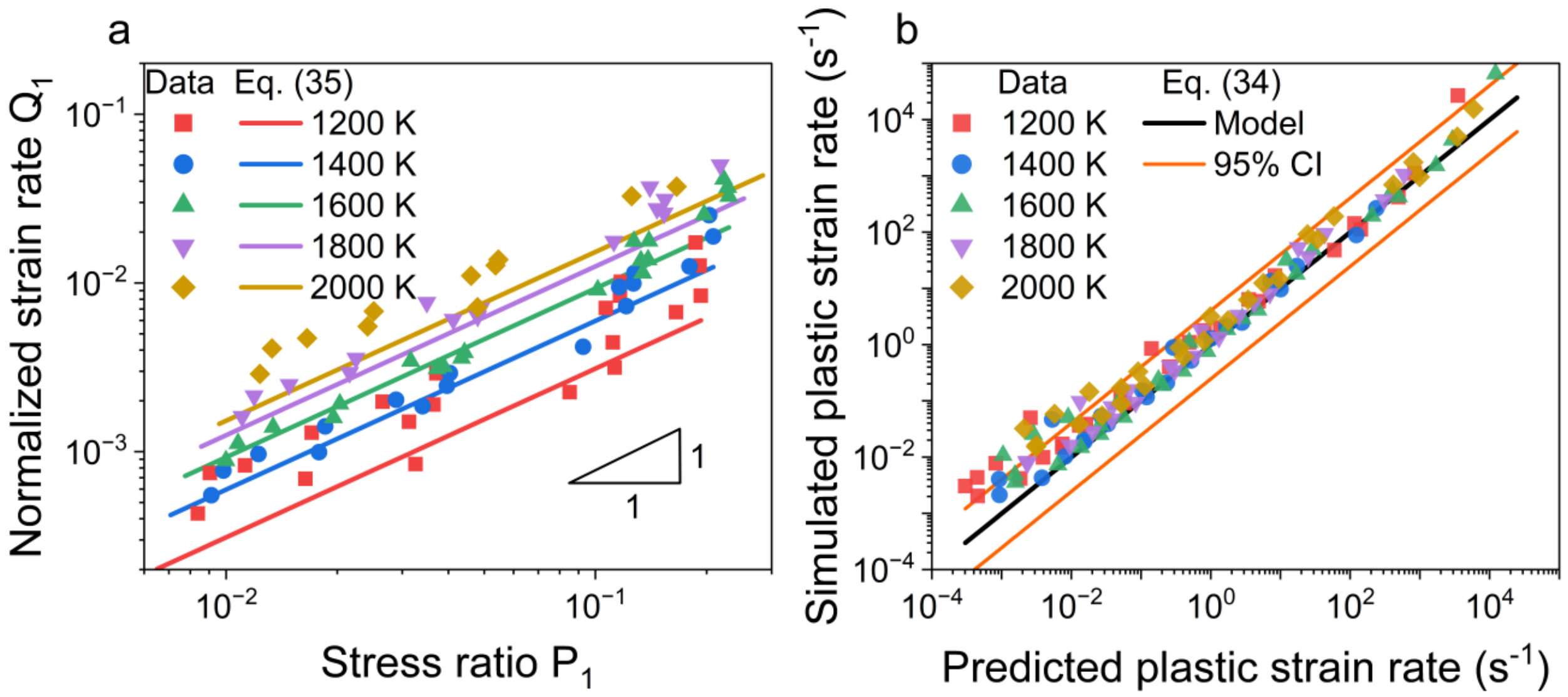

Fig. 6. Creep behavior under low-stress conditions. (a) Dimensionless scaling of the normalized parameters compared with the theoretical fitting curve. (b) Direct comparison between absolute DDD simulation results and predictions from the physical constitutive model.

### *3.6. Anomalous thermal hardening at high strain rate*

At high applied stress, the plastic strain rate decreases with increasing temperature. This counterintuitive relationship manifests in macroscopic properties: at a constant plastic strain rate, the material's strength increases while temperature rises. This anomalous behavior can be explained by the temperature-dependent enhancement of the phonon drag coefficient. To describe this phenomenon, the classical power-law constitutive model is employed,

$$\dot{\varepsilon}^{\mathrm{p}} = \dot{\varepsilon}_3^{\mathrm{p}} \rho_{\mathrm{a}} b^2 \left( \frac{\tilde{\tau}}{s_{\mathrm{e}}^0} \right)^m \exp\left( -\frac{\Delta \tilde{G}}{k_{\mathrm{B}} T} \right), \tag{36}$$

where $m$ is the power-law exponent, $\dot{\varepsilon}_3^{\mathrm{p}}$ is the reference strain rate, and $\Delta \tilde{G}$ is the apparent activation energy. Introduce the dimensionless stress ratio and plastic strain rate,

$$P_3 = \frac{\tilde{\tau}}{s_{\mathrm{e}}^0}, \quad Q_3 = \frac{\dot{\varepsilon}^{\mathrm{p}}}{\dot{\varepsilon}_3^{\mathrm{p}} \rho_a b^2}, \tag{37}$$

Eq. (36) can be expressed as:

$$\ln Q_3 = m \ln P_3 - \frac{\Delta \tilde{G}}{k_{\mathrm{B}} T} \tag{38}$$

To investigate the plasticity behavior under extreme loading conditions, Fig. 7(a) presents the proposed constitutive model across a wide spectrum of temperatures in the high-stress regime. By fixing a representative dislocation density, Fig. 7(b) further elucidates the coupled evolution of the applied resolved shear stress, plastic strain rate, and the apparent activation enthalpy. The most striking feature emerges in the dynamic response: as the strain rate increases, the apparent activation energy exhibits a monotonic decrease, eventually undergoing a distinct sign reversal (from positive to negative) at a critical strain rate threshold of approximately $3.5\times10^5$ $\mathrm{s}^{-1}$. This zero-crossing mathematically dictates a fundamental shift in the rate-limiting deformation mechanism. In the positive regime, plasticity is governed by conventional thermal activation, leading to thermal softening. Conversely, the negative activation energy regime signifies the dominance of phonon drag; here, elevated temperatures intensify lattice vibrations, thereby anomalously increasing the viscous drag on mobile dislocations. This framework not only captures the transition from thermal softening to anomalous thermal hardening but also achieves quantitative agreement with established high-strain-rate experimental observations (Borasi et al., 2026; Borasi and Schuh, 2025; Dowding and Schuh, 2024; Dowding and Schuh, 2026), providing a rigorous mechanistic basis for dynamic strengthening in BCC metals.

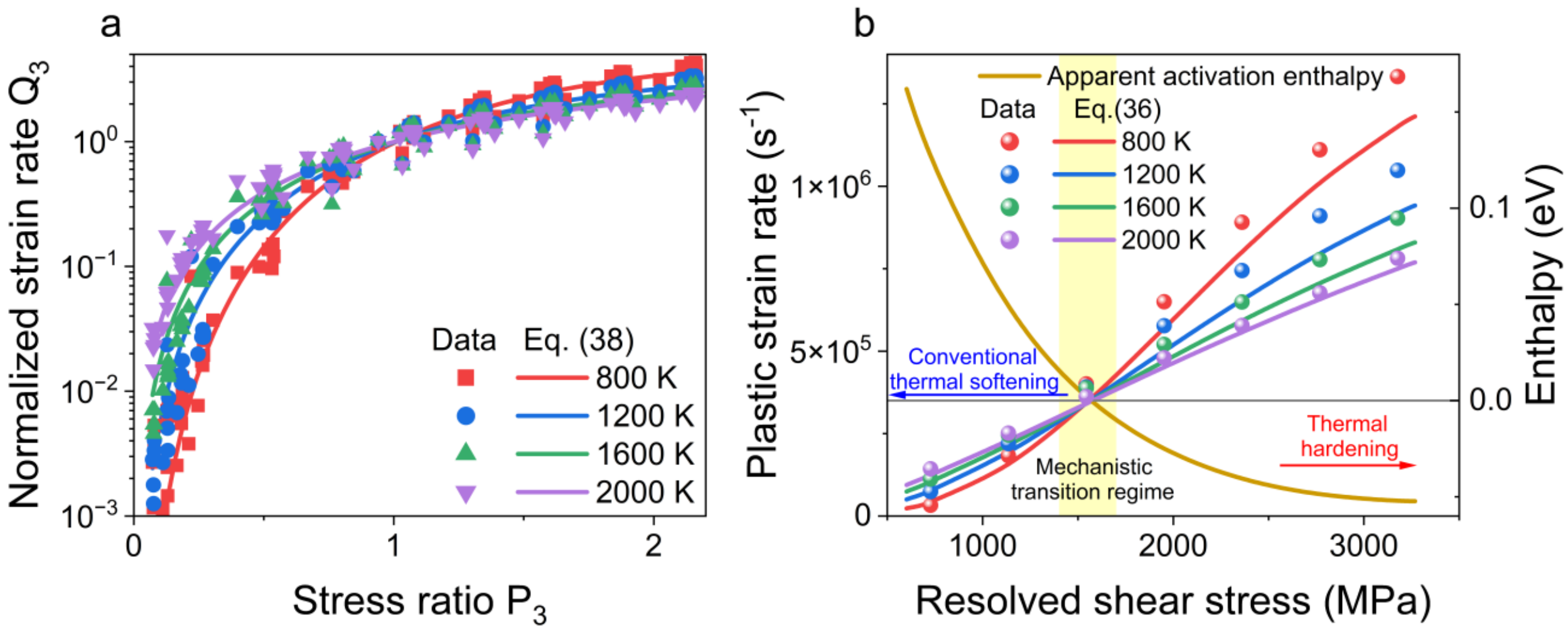

Fig. 7. Dynamic plasticity response and mechanism transition in the high-stress regime. (a) Model validation and data fitting of the proposed constitutive equation across a wide spectrum of temperatures under high-stress conditions. (b) The coupled evolution of the macroscopic plastic strain rate and the apparent activation enthalpy as a function of the resolved shear stress, evaluated at a representative constant dislocation density.

**Table 2** Parameters of strain rate and temperature dependent plasticity model

| | | | |
|---|---|---|---|
| $\dot{\varepsilon}_1^p\left(\mathrm{s}^{-1}\right)$ | $\dot{\varepsilon}_2^p\left(\mathrm{s}^{-1}\right)$ | $n$ | $\tilde{V}\left(\mathrm{b}^3\right)$ |
| $1.5299\times10^{11}$ | $2.4122\times10^{11}$ | 0.92779 | 1.6440 |
| $s_\mathrm{e}^0\left(\mathrm{MPa}\right)$ | $h\left(\mathrm{m}^{-3}\right)$ | $\alpha$ | $\beta$ |
| 1507.9 | $2.8128\times10^{14}$ | 0.4857 | 0.3 |
| $\mu\left(\mathrm{GPa}\right)$ | $\Delta\tilde{H}_0\left(\mathrm{eV}\right)$ | $p$ | $q$ |
| 161 | 1.506 | 0.887 | 2.000 |
| $\tilde{C}\left(\mathrm{eV}\right)$ | $\Delta\tilde{E}\left(\mathrm{eV}\right)$ | $\dot{\varepsilon}_3^p\left(\mathrm{s}^{-1}\right)$ | $m$ |
| 0.0110 | 0.6442 | $3.8291\times10^{11}$ | 0.90398 |

## 4. Discussion

The present study provides a mechanism-based route for connecting dislocation mobility to the macroscopic plastic response of BCC W. A central outcome is the distinction between two different types of superposition that are often mixed in phenomenological plasticity models. Structural barriers, including the Peierls resistance, forest interactions, and the carrier limitation associated with dislocation starvation, reduce the effective resolved shear stress available for glide. These contributions therefore enter the constitutive description as resistance stresses. In contrast, thermally activated kink-pair motion and phonon-drag-controlled glide represent distinct kinetic pathways for producing plastic strain. Their competition is most naturally described at the strain-rate level. This separation clarifies why stress superposition is appropriate for quasi-static resistance mechanisms, whereas the transition between thermal activation and viscous glide requires an additive decomposition of kinetic strain-rate contributions.

This distinction is important for BCC metals because the rate-controlling mechanism changes strongly with stress, temperature, and dislocation density. At low applied stresses, the mobility of screw dislocations is limited by the activation barrier for kink-pair formation. In this regime, increasing temperature promotes barrier crossing and leads to conventional thermal softening. The inclusion of both forward and reverse thermally activated transitions is essential in this limit because it enforces zero net glide velocity in the absence of a mechanical driving force. This feature removes an inconsistency of forward-only activation laws and provides a physically consistent low-stress limit for continuum constitutive modeling.

As the applied stress increases, the activation barrier is progressively reduced and the plastic response evolves toward phonon-drag-controlled glide. In this regime, temperature affects the plastic strain rate in the opposite direction. Since the phonon drag coefficient increases with temperature, a higher stress is required to sustain the same dislocation velocity and, consequently, the same plastic strain rate. The resulting increase in flow stress with temperature is not a contradiction of thermal activation theory; rather, it reflects a change in the dominant kinetic resistance from barrier crossing to lattice dissipation. The apparent negative activation energy extracted at high strain rates should therefore be interpreted as an effective descriptor of the macroscopic flow rule, not as a negative microscopic energy barrier. It marks the condition under which the temperature dependence of phonon drag outweighs the softening caused by thermally assisted glide.

The dislocation-density dependence further demonstrates that plastic strength cannot be determined by forest hardening alone. At very low dislocation densities, the scarcity of mobile carriers requires each dislocation to move at a much higher velocity in order to maintain a prescribed plastic strain rate. This produces an apparent strengthening associated with dislocation starvation. With increasing density, the availability of mobile dislocations initially facilitates plastic flow, whereas at high density the increasing frequency of dislocation interactions restores classical forest hardening. The resulting non-monotonic dependence of the critical stress on dislocation density provides a direct mechanistic link between microstructural state and macroscopic strength.

Several limitations remain. The current simulations focus on pure W and idealized initial dislocation configurations. Extension to refractory alloys will require additional resistance terms associated with solute-dislocation interactions and chemical disorder. In addition, the present model focuses on glide-controlled plasticity and does not explicitly account for non-Schmid effects associated with non-glide stress components. It also omits climb, cross-slip, diffusion-assisted recovery, or long-term dislocation-density evolution, all of which may become important during sustained high-temperature deformation. Future work should therefore couple the present mobility law with microstructure evolution models to improve its predictive capability for alloyed BCC systems and long-term service conditions.

## 5. Conclusion

In this paper, we derived a temperature- and strain-rate-dependent constitutive framework for BCC W by combining dislocation mobility laws with three-dimensional DDD simulations. The screw dislocation mobility law incorporates forward and reverse thermally activated glide, a smooth transition to phonon-drag-controlled motion, and a high-velocity correction. This formulation gives a physically consistent low-stress limit and avoids an artificial discontinuity between thermally activated and viscous glide regimes. The key findings are summarized below:

DDD simulations over a broad range of temperatures, applied stresses, and initial dislocation densities reveal a strong coupling among thermal activation, phonon drag, dislocation starvation, and forest hardening. The critical stress exhibits a non-monotonic dependence on dislocation density, decreasing in the carrier-limited regime and increasing again when forest interactions become dominant. This behavior shows that the density dependence of strength in BCC metals is governed by the competition between the availability of mobile dislocations and their mutual interactions.

A unified constitutive model is established by distinguishing structural resistance from kinetic strain-rate contributions. The Peierls stress, forest hardening, and dislocation-starvation resistance enter through an effective critical resistance, whereas thermal activation and phonon drag contribute as competing kinetic mechanisms at the strain-rate level. This separation enables the model to reproduce both low-stress thermally activated flow and high-stress phonon-drag-controlled plasticity.

The model also explains the transition from conventional thermal softening to anomalous thermal hardening at high strain rates. At low stresses, increasing temperature accelerates thermally activated screw-dislocation glide. At high stresses, however, the temperature-dependent increase in phonon drag reduces the dislocation velocity at a given stress, so that a higher flow stress is required to maintain the same plastic strain rate. The resulting sign change in the apparent activation energy identifies the crossover between these two rate-controlling mechanisms.

## Competing interests

The authors declare that they have no known competing financial interests or personal relationships that could have appeared to influence the work reported in this paper.

## Acknowledgments

Y.Z. acknowledges financial support from National Natural Science Foundation of China (Nos. 12588301 and 12472103), National Key Laboratory of Computational Physics (Nos. SYSM-2006-WDZC-06 and SYSZ-2025-ZZ-01) and Science Challenge Project (No.TZ2025016). The authors gratefully acknowledge the computational resources provided by the Computing Centers in Xi'an.

## Author contributions

Jiarui Zhang: Writing – original draft, Visualization, Validation, Software, Methodology, Formal analysis, Conceptualization. Yin Zhang: Writing – review & editing, Software, Investigation, Supervision, Project administration, Funding acquisition, Conceptualization.

## Appendix A. Algorithmic implementation and topological rules in DDD

To ensure the fidelity and numerical stability of the large-scale DDD simulations, the proposed nonlinear mobility law was integrated into the robust time-integration and topological framework of the ParaDiS code. The evolution of the complex 3D dislocation networks is governed by a rigorous interplay between the physical mobility laws and geometric discretization algorithms. While cross-slip is a known mechanism in BCC metals, it was not explicitly included as a stochastic event or an additional degree of freedom in the present DDD simulations.

**Integration algorithm and velocity calculation.** The simulation advances through a time integration scheme where the motion of dislocations is considered to be in the over-damped regime. In this regime, inertia is ignored, and the nodal forces determine instantaneous velocities. We modified the core velocity calculation subroutine (MobilityLaw_BCC_0.c) to implement our specific mobility model.

For a given dislocation network, the driving force on the nodes is balanced by the drag force. The nodal velocities are obtained by establishing a relationship between the drag coefficient matrix and the nodal forces. Specifically, the velocity $v_i$ of node $i$ is determined by the damping matrix formulated from its neighbouring nodes $j$, along with the node force:

$$\sum_j \frac{\left\| l_{ij} \right\|}{2} B_{ij} \cdot \mathbf{v}_j = \mathbf{F}_i, \tag{39}$$

where $B_{ij}$ represents the drag coefficient matrix, $\mathbf{F}_i$ is the nodal force, and $\left\| l_{ij} \right\|$ denotes the length of the dislocation line segment connecting node $i$ and its neighbouring node $j$.

In our implementation, the drag coefficient matrix $B_{ij}$ is constructed by explicitly calculating the mobility coefficients for edge and screw dislocation components based on the nodal forces:

$$B_{ij} = B_{\mathrm{g}}\left(\mathbf{m}\otimes\mathbf{m}\right) + B_{\mathrm{c}}\left(\mathbf{n}\otimes\mathbf{n}\right), \tag{40}$$

where **t** is the dislocation line vector, **b** represents the Burgers vector, **m** is the normal vector to the dislocation climb plane, and **n** is the normal vector to the dislocation glide plane, and

$$\begin{aligned} B_{\mathrm{g}} &= \frac{1}{\|\mathbf{b}\|}\left[B_{\mathrm{eg}}^{-2}\|\mathbf{b}\times\mathbf{t}\|^2 + B_{\mathrm{s}}^{-2}\|\mathbf{b}\cdot\mathbf{t}\|^2\right]^{-\frac{1}{2}}, \\ B_{\mathrm{c}} &= \frac{1}{\|\mathbf{b}\|}\left[B_{\mathrm{ec}}^{2}\|\mathbf{b}\times\mathbf{t}\|^2 + B_{\mathrm{s}}^{2}\|\mathbf{b}\cdot\mathbf{t}\|^2\right]^{\frac{1}{2}}, \end{aligned} \tag{41}$$

here, $B_{\mathrm{eg}}$ denotes the glide mobility coefficient for edge dislocations, $B_{\mathrm{ec}}$ is the climb mobility coefficient for edge dislocations, and $B_{\mathrm{s}}$ represents the mobility coefficient for screw dislocations. Consequently, the dislocation velocity is given by:

$$\mathbf{v}_i = \left(\sum_j \frac{\|l_{ij}\|}{2} B_{ij}\right)^{-1} \cdot \mathbf{F}_i. \tag{42}$$

In our formulation, we establish that $B_{\mathrm{s}}$ depends on both stress and temperature, whereas $B_{\mathrm{eg}}$ depends on temperature. Because the motion of discretized nodes along the dislocation line does not alter the dislocation configuration, such tangential movement should not encounter any physical resistance. Consequently, we consider the mobility to be governed by the stress acting on the plane perpendicular to the dislocation line. Therefore, we define:

$$B_{\mathrm{eg}} = B_{\mathrm{eg}}(\overline{\mathbf{F}}_i, T), \quad B_{\mathrm{s}} = B_{\mathrm{s}}(\overline{\mathbf{F}}_i), \tag{43}$$

where

$$\overline{\mathbf{F}}_i = \left(\mathbf{I}_2 - \mathbf{t}\otimes\mathbf{t}\right)\mathbf{F}_i / \|l_{ij}\|. \tag{44}$$

The nodal velocities are then directly computed from the overall damping matrix. Notably, we utilized a direct calculation approach rather than an iterative solver to determine the velocities from the forces.

**Network remeshing.** Continuous dislocation lines are discretized into piecewise straight segments connected by nodes. To maintain a physically accurate representation of the network geometry while optimizing computational cost, dynamic remeshing is continuously executed. This involves adding or deleting nodes based on geometric criteria. Specifically, nodes are added when segments become too long (exceeding a maximum length, $l_{\max}$) or to better resolve areas of high local curvature. Conversely, nodes are deleted if a segment becomes too short (falling below a minimum length, $l_{\min}$) to prevent unnecessary computational burden.

**Topological Operations.** The formation of complex entanglements relies on the collision algorithm. Topological changes, such as recombination or junction formation, are essential for handling the interactions of dislocation lines. When two distinct, non-connected dislocation segments approach each other within a specified collision distance parameter, they are considered to be in contact. The code then introduces new nodes at the points of closest approach on the colliding segments. These new nodes are subsequently merged to form a single node. To enable dislocation reactions, this resulting multi-arm node can split into two new nodes. The choice of how the node splits is determined by selecting the topological outcome that maximizes the local rate of energy dissipation, allowing the network to evolve towards a minimum of free energy.

## Appendix B. Determination of steady-state plastic strain rates

To accurately quantify macroscopic plastic flow characteristics under varying initial dislocation densities, applied stresses, and temperatures, an optimized fitting approach was employed to extract the steady-state plastic strain rate. This approach effectively mitigates the interference from transient structural evolution and high-frequency numerical noise during the initial stages of DDD simulations.

Directly time-averaging the plastic strain rate, $\dot{\varepsilon}^{\mathrm{p}}$, introduces substantial error due to the inevitable transient activation of dislocation sources and configurational relaxation upon initial constant-stress loading. To address this limitation, representative steady-state parameters were extracted based on the integral form of the classic Orowan equation:

$$\varepsilon^{\mathrm{p}}\left(t\right)=\int_0^t \dot{\varepsilon}^{\mathrm{p}}\left(\tau\right)\mathrm{d}\tau=\int_0^t \rho\left(\tau\right)b\bar{v}\mathrm{d}\tau. \tag{45}$$

Assuming a constant average dislocation velocity, $\bar{v}$, during the quasi-steady-state flow regime, the plastic strain, $\varepsilon^{\mathrm{p}}$, exhibits a strictly linear relationship with the time integral of the dislocation density, $\int \rho \mathrm{d}t$. The slope of this linear correlation is defined as the evolution constant, $k = b\bar{v}$. To objectively and automatically identify the steady-state flow regime within complex time-series data, a maximum continuous linear segment search algorithm was implemented. This algorithm performs a sliding-window search across the integral domain, $\left(\int \rho \mathrm{d}t, \varepsilon^{\mathrm{p}}\right)$. The objective function aims to locate the longest continuous data segment that satisfies a rigorous linearity criterion (coefficient of determination, $R^2 \geq 0.98$). Upon extracting the optimal slope, $k$, for a given configuration under specific loading conditions, the corresponding characteristic plastic strain rate is defined as:

$$\dot{\varepsilon}^{\mathrm{p}}_{\mathrm{steady}} = k \cdot \rho_0 \tag{46}$$

where $\rho_0$ represents the initial dislocation density. Mathematically, this methodology effectively decouples the flow rate from the transient interference caused by the dynamic multiplication and annihilation of dislocations during the simulation.

When the average dislocation velocity falls below a threshold of $10^{-14}$ m/s, the velocity is considered to be dominated by numerical fluctuations and the corresponding data points are excluded from the analysis. If the network reached a stationary configuration before a segment satisfied these criteria, the trajectory was classified as excluded.

## Appendix C. Statistical robustness of DDD simulation results

Fig. A1 summarizes the plastic flow responses for the simulated combinations of temperature, applied stress and initial dislocation density. For each parameter combination, six loading cases were initiated from the same relaxed dislocation configuration. These comprised uniaxial tension and compression along each of the three orthogonal <100> directions. The initial configuration was relaxed without external loading, but residual configurational evolution or numerical self-deformation could persist during subsequent loading. Accordingly, tensile and compressive responses in each direction were averaged to reduce the influence of such signed residual deformation. Averaging over the three directions further reduced sensitivity to the selected loading direction. For a complete set of six accepted loading cases, the reported response was their equally weighted arithmetic mean. Error bars denote the standard deviation across the six case-specific responses.

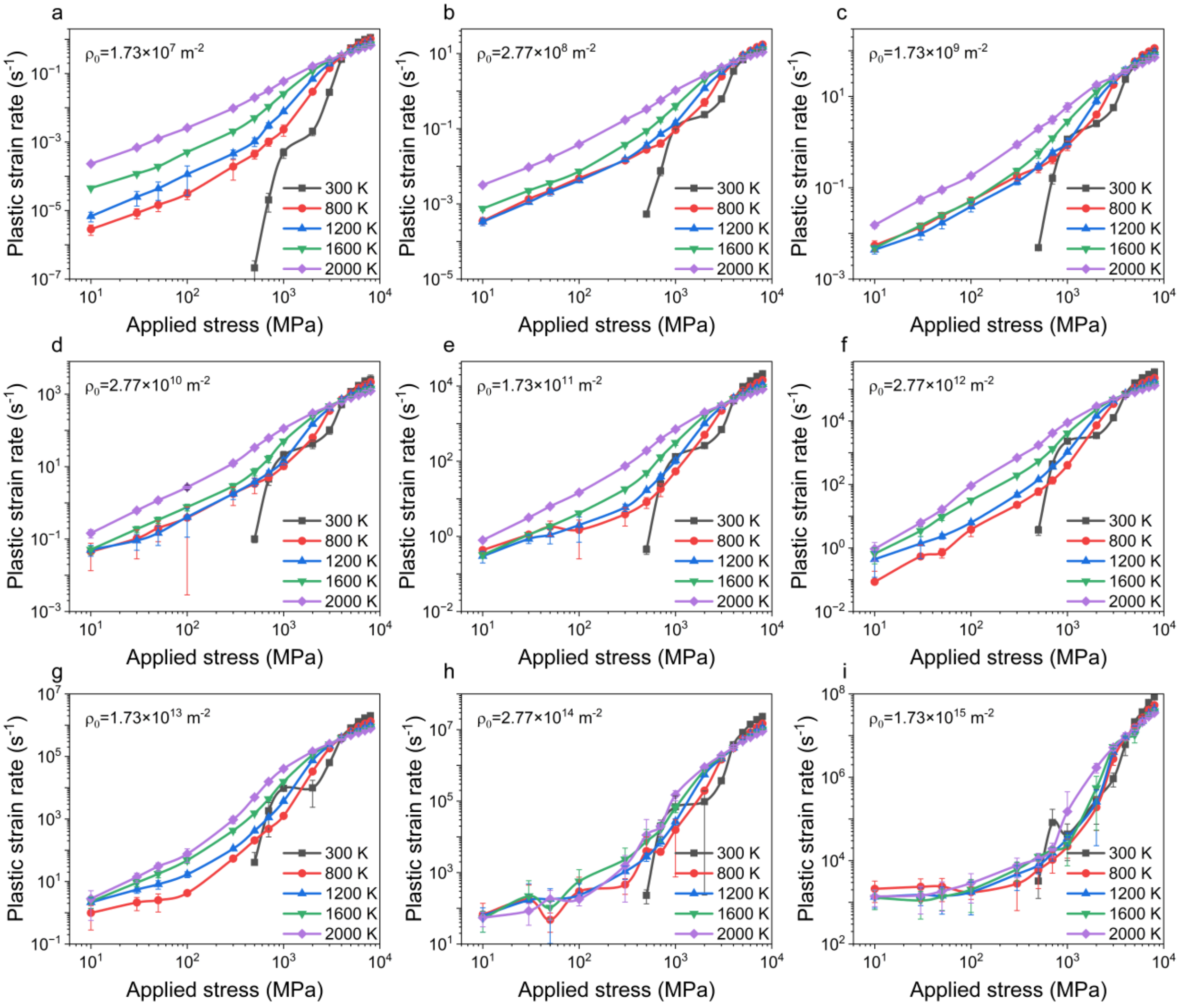

Fig. A1. The simulation results under different initial dislocation densities. (a)-(i) correspond to initial dislocation densities ranging from $10^7$ to $10^{15}$ $m^{-2}$. For complete six-case sets, each data point is the mean and each error bar the standard deviation of the six loading-case responses defined in Appendix C.

# References

Al'shitz, V.A., Indenbom, V.L., 1975. Dynamic dragging of dislocations. Phys.-Usp. 18, 1-20.

Argon, A., 2007. Strengthening Mechanisms in Crystal Plasticity. Oxford University Press.

Arsenlis, A., Cai, W., Tang, M., Rhee, M., Oppelstrup, T., Hommes, G., Pierce, T.G., Bulatov, V.V., 2007. Enabling strain hardening simulations with dislocation dynamics. Modell. Simul. Mater. Sci. Eng. 15, 553-595.
Ashby, M.F., Jones, D.R.H., 2012. Engineering Materials 1. Butterworth-Heinemann.
Borasi, L., Kooi, S.E., Schuh, C.A., 2026. Crossing from thermally activated to drag-controlled plasticity in mild steel as strain rate increases. Acta Mater. 310, 122120.
Borasi, L., Schuh, C.A., 2025. Self-consistent hardness measurements spanning eleven decades of strain rate on a single material surface. Nat. Commun. 16, 6148.
Bulatov, V., Cai, W., 2006. Computer Simulations of Dislocations. Oxford University Press.
Bulatov, V.V., Hsiung, L.L., Tang, M., Arsenlis, A., Bartelt, M.C., Cai, W., Florando, J.N., Hiratani, M., Rhee, M., Hommes, G., Pierce, T.G., de la Rubia, T.D., 2006. Dislocation multi-junctions and strain hardening. Nature 440, 1174-1178.
Cereceda, D., Diehl, M., Roters, F., Raabe, D., Perlado, J.M., Marian, J., 2016. Unraveling the temperature dependence of the yield strength in single-crystal tungsten using atomistically-informed crystal plasticity calculations. Int. J. Plast. 78, 242-265.
Chang, J., Cai, W., Bulatov, V.V., Yip, S., 2002. Molecular dynamics simulations of motion of edge and screw dislocations in a metal. Comput. Mater. Sci. 23, 111-115.
Cottrell, A.H., 2006. The Nabarro equation for thermally activated plastic glide. Philos. Mag. 86, 3811-3817.
Curtin, W.A., Ronald, E.M., 2003. Atomistic/continuum coupling in computational materials science. Modell. Simul. Mater. Sci. Eng. 11, R33.
Devincre, B., Madec, R., Monnet, G., Queyreau, S., Gatti, R., Kubin, L., 2011. Modeling crystal plasticity with dislocation dynamics simulations: The “microMegas” code, in: Samuel, F., Anne, P., Olivier, T. (Eds.), Mechanics of Nano-objects. Presses des Mines, Paris, pp. 81-100.
Domain, C., Monnet, G., 2005. Simulation of screw dislocation motion in iron by molecular dynamics simulations. Phys. Rev. Lett. 95, 215506.
Dowding, I., Schuh, C.A., 2024. Metals strengthen with increasing temperature at extreme strain rates. Nature 630, 91-95.
Dowding, I., Schuh, C.A., 2026. At extreme strain rates, pure metals thermally harden while alloys thermally soften. Phys. Rev. Lett. 136, 076101.
El-Awady, J.A., 2015. Unravelling the physics of size-dependent dislocation-mediated plasticity. Nat. Commun. 6, 5926.
El Ters, P., Shehadeh, M.A., 2019. Modeling the temperature and high strain rate sensitivity in BCC iron: Atomistically informed multiscale dislocation dynamics simulations. Int. J. Plast. 112, 257-277.
Faisal, A.H.M., Weinberger, C.R., 2019. Exponent for the power-law relation between activation energy for dislocation nucleation and applied stress. Phys. Rev. Mater. 3, 103601.
Fan, H., Wang, Q., El-Awady, J.A., Raabe, D., Zaiser, M., 2021. Strain rate dependency of dislocation plasticity. Nat. Commun. 12, 1845.
Ghoniem, N.M., Tong, S.H., Sun, L.Z., 2000. Parametric dislocation dynamics: A thermodynamics-based approach to investigations of mesoscopic plastic deformation. Phys. Rev. B 61, 913-927.
Giessen, E.V.d., Needleman, A., 1995. Discrete dislocation plasticity: a simple planar model. Modell. Simul. Mater. Sci. Eng. 3, 689-735.
Gilbert, M.R., Queyreau, S., Marian, J., 2011. Stress and temperature dependence of screw dislocation mobility in α-Fe by molecular dynamics. Phys. Rev. B 84, 174103.
Gillis, P.P., Gilman, J.J., Taylor, J.W., 1969. Stress dependences of dislocation velocities. Philos. Mag. 20, 279-289.
Gurrutxaga-Lerma, B., Verschueren, J., Sutton, A.P., Dini, D., 2020. The mechanics and physics of high-speed dislocations: a critical review. Int. Mater. Rev. 66, 215-255.
Hull, D., Bacon, D.J., 2011. Introduction to Dislocations. Butterworth-Heinemann.

Johnston, W.G., Gilman, J.J., 1959. Dislocation velocities, dislocation densities, and plastic flow in lithium fluoride crystals. J. Appl. Phys. 30, 129-144.
Kocks, U.F., Argon, A., Ashby, M., 1975. Thermodynamics and kinetics of slip. Pergamon.
Kocks, U.F., Mecking, H., 2003. Physics and phenomenology of strain hardening: the FCC case. Prog. Mater. Sci. 48, 171-273.
Kubin, L.P., Canova, G., Condat, M., Devincre, B., Pontikis, V., Bréchet, Y., 1992. Dislocation microstructures and plastic flow: A 3D simulation. Solid State Phenom. 23-24, 455-472.
Kuksin, A.Y., Yanilkin, A.V., 2013. Atomistic simulation of the motion of dislocations in metals under phonon drag conditions. Phys. Solid State 55, 1010-1019.
Kumagai, T., Takahashi, A., Takahashi, K., Nomoto, A., 2020. Velocity of mixed dislocations in body centered cubic iron studied by classical molecular dynamics calculations. Comput. Mater. Sci. 180, 109721.
Li, Z., Liu, Z., Zhuang, Z., Cui, Y., 2021. Temperature dependent deformation localization in irradiated tungsten. Int. J. Plast. 146, 103077.
Monavari, M., Zaiser, M., 2018. Annihilation and sources in continuum dislocation dynamics. Mater. Theory 2, 3.
Nabarro, F.R.N., 1997. Fifty-year study of the Peierls-Nabarro stress. Mat. Sci. Eng. A 234-236, 67-76.
Narayanan, S., McDowell, D.L., Zhu, T., 2014. Crystal plasticity model for BCC iron atomistically informed by kinetics of correlated kinkpair nucleation on screw dislocation. J. Mech. Phys. Solids 65, 54-68.
Po, G., Cui, Y., Rivera, D., Cereceda, D., Swinburne, T.D., Marian, J., Ghoniem, N., 2016. A phenomenological dislocation mobility law for bcc metals. Acta Mater. 119, 123-135.
Roters, F., Eisenlohr, P., Hantcherli, L., Tjahjanto, D.D., Bieler, T.R., Raabe, D., 2010. Overview of constitutive laws, kinematics, homogenization and multiscale methods in crystal plasticity finite-element modeling: Theory, experiments, applications. Acta Mater. 58, 1152-1211.
Saroukhani, S., Warner, D.H., 2017. Investigating dislocation motion through a field of solutes with atomistic simulations and reaction rate theory. Acta Mater. 128, 77-86.
Škoro, G.P., Bennett, J.R.J., Edgecock, T.R., Gray, S.A., McFarland, A.J., Booth, C.N., Rodgers, K.J., Back, J.J., 2011. Dynamic Young's moduli of tungsten and tantalum at high temperature and stress. J. Nucl. Mater. 409, 40-46.
Srivastava, K., Weygand, D., Caillard, D., Gumbsch, P., 2020. Repulsion leads to coupled dislocation motion and extended work hardening in bcc metals. Nat. Commun. 11, 5098.
Starikov, S., Tseplyaev, V., 2020. Two-scale simulation of plasticity in molybdenum: Combination of atomistic simulation and dislocation dynamics with non-linear mobility function. Comput. Mater. Sci. 179, 109585.
Stein, D.F., Low, J.R., 1960. Mobility of edge dislocations in silicon-iron crystals. J. Appl. Phys. 31, 362-369.
Stukowski, A., Cereceda, D., Swinburne, T.D., Marian, J., 2015. Thermally-activated non-Schmid glide of screw dislocations in W using atomistically-informed kinetic Monte Carlo simulations. Int. J. Plast. 65, 108-130.
Weinberger, C.R., Jennings, A.T., Kang, K., Greer, J.R., 2012. Atomistic simulations and continuum modeling of dislocation nucleation and strength in gold nanowires. J. Mech. Phys. Solids 60, 84-103.
Zaiser, M., Nikitas, N., Hochrainer, T., Aifantis, E.C., 2007. Modelling size effects using 3D density-based dislocation dynamics. Philos. Mag. 87, 1283-1306.
Zbib, H.M., Diaz de la Rubia, T., 2002. A multiscale model of plasticity. Int. J. Plast. 18, 1133-1163.
Zhu, T., Li, J., Samanta, A., Leach, A., Gall, K., 2008. Temperature and strain-rate dependence of surface dislocation nucleation. Phys. Rev. Lett. 100, 025502.
Zhu, T., Li, J., Yip, S., 2013. Atomistic reaction pathway sampling: The nudged elastic band method and nanomechanics applications, Nano Cell Mech.: Fundament. Front., pp. 311-338.